\documentclass[conference]{IEEEtran}
\IEEEoverridecommandlockouts
\usepackage{cite}
\usepackage{amsmath,amssymb,amsfonts}
\usepackage{algorithmic}
\usepackage{graphicx}
\usepackage{textcomp}
\usepackage{xcolor}
\def\BibTeX{{\rm B\kern-.05em{\sc i\kern-.025em b}\kern-.08em
    T\kern-.1667em\lower.7ex\hbox{E}\kern-.125emX}}
\usepackage{amsthm}
\usepackage{ascmac}
\usepackage{enumitem}
    
\newtheorem{Theorem}{Theorem}
\newtheorem{Definition}{Definition}

\newtheorem{Lemma}{Lemma}

\newtheorem{Corollary}{Corollary}
\newcommand{\Var}{\mathrm{Var}}

\begin{document}

\title{Secret {Key-based} Authentication With Passive Eavesdropper for Scalar Gaussian Sources}
\vspace{0mm}
\author{\IEEEauthorblockN{ Vamoua Yachongka~~~~~~~~~~~~~~~~ Hideki Yagi~~~~~~~~~~~~~~~~Yasutada Oohama}
\IEEEauthorblockA{\textit{Dept. Network and Computer Engineering},
\textit{The University of Electro-Communications}, Chofu, Tokyo, Japan\\
\{va.yachonka, h.yagi, oohama\}@uec.ac.jp}
}

\maketitle

\begin{abstract}
We analyze the fundamental trade-off of secret {key-based} authentication systems in the presence of an {eavesdropper} for correlated Gaussian sources. A complete characterization of {the} trade-off among secret-key, storage, and privacy-leakage rates of both generated and chosen secret models is provided. {One} of the main contributions {is revealing} that unlike the known results {for} discrete sources, there is no need for the second auxiliary random variable in characterizing the capacity regions for {the} Gaussian {cases}. {In addition, it is shown} that the {strong secrecy for} secrecy-leakage of the systems can be achieved by an information-spectrum approach, and the parametric expressions ({computable} forms) of the capacity regions are also derived.
\end{abstract}


\section{Introduction}

Secret key-based authentication (SKA) systems are generally designed to perform private authentication of users based on secret keys, usually generated from biometric identifiers \cite{jian2009} or physical unclonable functions \cite{bohm2012}. In recent years, there has been a bunch of literature focusing on investigating the fundamental limits of SKA systems from information-theoretic perspectives. {In} the analysis of the {SKA} systems, {a new condition} called privacy constraint is {added to} the problem formulations of the well-known secret-key agreement {{(the source model with one-way communication only)}} discussed in, e.g., \cite{ac1993}--\hspace{-0.1mm}\cite{wataoha2010}. Therefore, many existing tools used for solving the key agreement problems are quite useful {to characterize} the capacity regions of the {SKA} systems as well. 

The seminal works \cite{itw3} and \cite{lhp} independently investigated the trade-off relation between security and privacy-leakage in the SKA systems. {Particularly},
in \cite{itw3}, eight different systems were taken into consideration, but among them the generated secret (GS) and chosen secret (CS) models are two {major} models closely related to real-life applications and frequently analyzed in the researches taking place later on.
Some extensions of the work \cite{itw3} for GS and CS models {considering} a storage constraint and {user identification can be found in, e.g., \cite{KY}--\hspace{-0.1mm}\cite{gunlu2019} and \cite{itw}--\hspace{-0.1mm}\cite{vy2}, respectively}.


{An} SKA system in which an {eavesdropper} can observe both the helper {data} and correlated {side information} of the identified sequence was introduced in \cite{kc2015}.
More specifically, the GS model {was} discussed in two scenarios; passive and active {eavesdropper} scenarios. In {the} passive case, the {eavesdropper} is interested in knowing the biometric identifier and the secret key based on the available information in {his/her} hand. On the other hand, the active {eavesdropper} {tries} to cheat the system or decode the genuine secret key with a mock sequence generated by {his/her} own data. Basically, in this setup, the privacy requirement is more stringent than the one seen in \cite{itw3}, in which the {eavesdropper} only {has} the knowledge of the helper. The work \cite{kc2015} was further {extended} in \cite{gksc2018}, \cite{gs2021} to incorporate noisy enrollment and a cost-constrained action at the decoder. The {capacity characterizations} of {each} paper were derived via two auxiliary random variables (RVs) for discrete memoryless sources as seen in \cite{cn2000}. However, in practical applications, the biometric signal is usually represented in continuous forms.

Moreover, in information-theoretic security, the weak secrecy and the strong secrecy are commonly defined as metrics to assess the leakage of sensitive data, e.g, secret key used for private authentication.
In \cite{kc2015}, \cite{gksc2018}, the secrecy-leakage of the systems is evaluated under a weak secrecy criterion, where the secret-key information is allowed to leak to the {eavesdropper} in sub-linear order of the block length. From the security point of views, this is not preferable for the reason that the information leaked to the {eavesdropper} might grow unbounded with the block length \cite{BB}.
A tighter security notion imposed on the secrecy-leakage is seen in \cite{gunlu2019}, \cite{gs2021}. In the papers, {an} SKA system, in which the noisy identifiers and the identified sequences are observed through a broadcast channel \cite[Chapter 8]{GK}, is analyzed under a strong secrecy criterion for secrecy-leakage, where the amount of information leaked to the {eavesdropper} is demanded to be negligible regardless of the block length. However, the analyzing technique is different from the one {adopted} in this paper.





Motivated by these {essential factors}, for the same model in \cite{kc2015}, we enhance the security criterion to the strong secrecy for secrecy-leakage, and characterize the capacity regions {of secret-key, storage, and privacy-leakage rates} for Gaussian sources {from information-spectrum perspectives} \cite{han2003}. We solely focus on the passive {eavesdropper} case, but deal with both GS and CS models. The main contributions of this paper are summarized as follows:
\begin{itemize}
    \item Show that unlike the results of discrete sources \cite{kc2015}--\hspace{-0.1mm}\cite{gs2021}, a single auxiliary RV suffices to characterize the capacity regions of both models for Gaussian sources. Another interesting {result} is when the correlation coefficient of {the channel to the decoder} is smaller than that of {the channel to the eavesdropper}, the capacity regions of the two models coincide; {the} optimal rates of the secret key and storage {become} zero, but the minimum value of the privacy-leakage rate {may} be positive.
    \item Apply a {privacy amplification technique} developed in \cite{wataoha2010} to prove that the strong {secrecy for} secrecy-leakage of the {SKA} systems is achievable.
    \item {Provide} complete parametric expressions of the capacity regions.
\end{itemize}

As special cases, the capacity {characterizations} derived in this paper coincide with the results of \cite{wataoha2010} for {a} large enough privacy-leakage rate and \cite{willems2009} for no consideration of {the} storage rate.
\vspace{0mm}
\section{System Models and Problem Formulations}
\subsection{System Models}

\begin{figure}[!t]
    \vspace{0mm}
    \centering
    \includegraphics[scale=0.53]{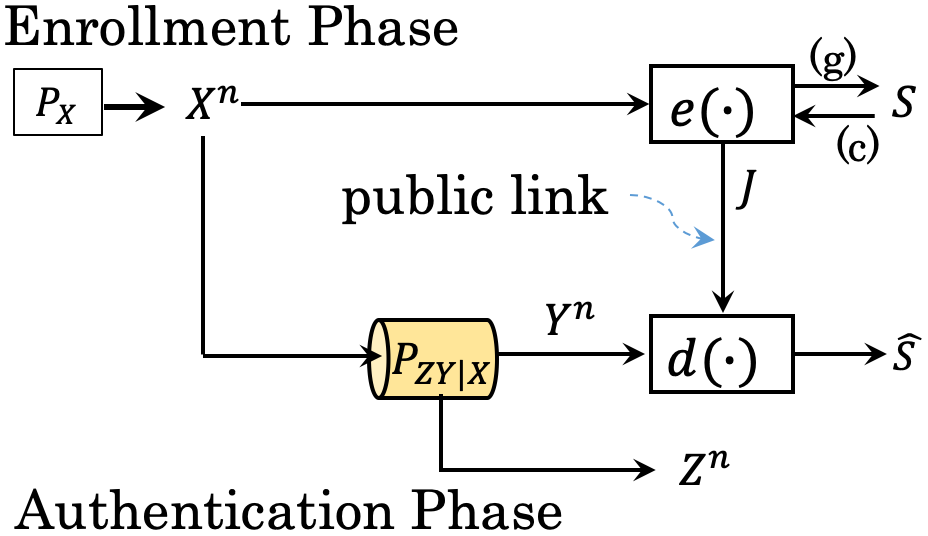}
    \vspace{0mm}
    \caption{System models}
    \label{fig:model}
    \vspace{0mm}
\end{figure}

We basically use standard notation in \cite{GK}. The data flow of GS and CS models is depicted in Fig.\ \ref{fig:model}. {Arrows} (g) and (c) {indicate} the directions of the secret key of the former and {latter} models.
Assume that the biometric source $X \sim \mathcal{N}(0,1)$. {The channel to the decoder $(X \rightarrow Y)$ and the channel to the eavesdropper $(X \rightarrow Z)$} are modeled as
\vspace{0mm}
\begin{align}
    Y = \rho_1X + N_y,~~~~~ Z = \rho_2X + N_2, \label{zyxn12}
\end{align}
where $|\rho_1|,|\rho_2| < 1$ are the correlation coefficients of each channel, $N_y \sim \mathcal{N}(0,1-\rho^2_1)$ and $N_2 \sim \mathcal{N}(0,1-\rho^2_2)$ are Gaussian RVs, and independent of each other and {of} other RVs. Let $\mathcal{S} = [1:M_S]$ and $\mathcal{J} = [1:M_J]$ be the sets of secret keys and helper data, respectively. $X^n$ and $(Y^n,Z^n)$ denote the biometric identifier generated from source $P_{X}$, the outputs of $X^n$ via the channel $P_{YZ|X}$, respectively. In GS model, observing $X^n$, the encoder $e$ generates {a} helper data $J \in \mathcal{J}$ and a secret key $S \in \mathcal{S}$; $(J,S) = e(X^n)$. $J$ is shared with the decoder via the noiseless public channel. Seeing $Y^n$, the decoder $d$ estimates {$S$} from $Y^n$ and the helper data $J$; $\widehat{S}=d(Z^n,J)$. In CS model, $S$ is chosen uniformly from $\mathcal{S}$ and independent of other RVs. The encoder makes the helper data by $J = e(X^n,S)$. For the decoder, $\hat{S} = d(Y^n,J)$. The {eavesdropper} has $(Z^n,J)$ and wants to learn about the biometric identifier $X^n$ and the secret key $S$.
\subsection{Problem Formulations}
In this section, we provide the formal definitions of GS and CS models.
First, we define the achievability definition of GS model.
\vspace{0mm}
\begin{Definition} \label{def1}
A tuple of secret-key, storage, and privacy-leakage rates $(R_S,R_J,R_L)\in \mathbb{R}^3_+$ is said to be achievable for GS model if for any $\delta > 0$ and large enough $n$ there exist pairs of encoders and decoders satisfying
\vspace{0mm}
\begin{align}
\Pr\{\widehat{S} \neq S\} &\leq  \delta,~~~~~~~~~~~~~~~\mathrm{(error ~probability)}\label{errorp} \\
H(S) &\geq n(R_S - \delta),~~~~\mathrm{(secret\text{-}key)} \label{secretk} \\
\log{M_J} &\leq n(R_J + \delta),~~~~\mathrm{(storage)} \label{storage} \\
I(S;J,Z^n) &\leq \delta,~~~~~~~~~~~~~~~~\mathrm{(secrecy\text{-}leakage)} \label{secrecy} \\
I(X^n;J,Z^n) &\leq n(R_L + \delta).~~~~~\mathrm{(privacy\text{-}leakage)} \label{privacy}
\end{align}
Also, $\mathcal{R}_G$ is defined as the closure of the set of all achievable rate tuples for GS model, called the capacity region.
\end{Definition}

The achievability definition of CS model is given below.
\vspace{0mm}
\begin{Definition} \label{def2}
A tuple of $(R_S,R_J,R_L)\in \mathbb{R}^3_+$ is said to be achievable for CS model if for any $\delta > 0$ and large enough $n$ there exist pairs of encoders and decoders satisfying all the requirements imposed in Definition \ref{def1}. Let $\mathcal{R}_C$ be the capacity region of CS model.
\end{Definition}
\vspace{0mm}
\section{Statement of Main Results}

Before {stating} our main {theorem}, we would like to {briefly} mention the background why the general { authentication (wiretap) channels} can always be scaled down into the degraded {versions for Gaussian sources}.

Using \cite[Lemma 6]{wataoha2010}, in the case where
\vspace{0mm}
\begin{align}
    \rho^2_1 > \rho^2_2, \label{1ge2}
\end{align}
by {setting}
$
    X'=X,~Y'=Y,~Z' = \frac{\rho_2}{\rho_1} Y' + N_z,
$
where $N_z \sim \mathcal{N}(0,1-\rho^2_2/\rho^2_1)$ is Gaussian RV and independent of other RVs, the marginal densities of $(X',Y')$ and $(X',Z')$ coincides with $(X,Y)$ and $(X,Z)$, respectively. {Contrary} to the above condition, when
\vspace{0mm}
\begin{align}
    \rho^2_1 \le \rho^2_2, \label{1le2}
\end{align}
for the RVs
$
    X'=X,~Z'=Z,~Y' = \frac{\rho_1}{\rho_2} Z',
$
it also follows that the marginal densities of $(X',Y')$ and $(X',Z')$ coincides with $(X,Y)$ and $(X,Z)$, respectively.


{Since the constraints \eqref{errorp}, \eqref{privacy}, and \eqref{secrecy} {depend only} on the marginal densities of $(X,Y)$ and $(X,Z)$,
it suffices to derive our main {theorem based on the joint sources $(X',Y',Z')$} instead of \eqref{zyxn12}. In the rest of discussions, we just use $(X,Y,Z)$ to represent $(X',Y',Z')$ for convenience.}
\vspace{0mm}
\begin{Theorem}\label{th1}
Under the condition of \eqref{1ge2}, {i.e., the Markov chain $X-Y-Z$ holds}, we have that
\vspace{0mm}
\begin{align}
\mathcal{R}_G = \cup_{P_{U|X}}\{(R_S,&R_J,R_L)\in \mathbb{R}^3_+:
R_S \leq I(Y;U|Z),\nonumber \\
&R_J \geq I(X;U|Y),\nonumber \\
&R_L \geq I(X;U|Y) + I(X;Z)\}, \label{theorem1} \\
\mathcal{R}_C = \cup_{P_{U|X}}\{(R_S,&R_J,R_L)\in \mathbb{R}^3_+:
R_S \leq I(Y;U|Z),\nonumber \\
&R_J \geq I(X;U|Z),\nonumber \\
&R_L \geq I(X;U|Y) + I(X;Z)\}. \label{theorem2}
\end{align}
For the case of \eqref{1le2}, {i.e., the Markov chain $X-Z-Y$ holds}, the capacity regions are characterized in the same form as
\vspace{0mm}
\begin{align}
\mathcal{R}_G = \mathcal{R}_C = \{(R_S,R_J,R_L):~&R_S = 0,~~R_J \geq 0, \nonumber \\
&R_L \geq I(X;Z)\}. \label{theorem3}
\end{align}
\qed
\end{Theorem}

A proof of Theorem \ref{th1} is given in Appendix A.
Note that the above regions are uncomputable since the cardinality of auxiliary RV $U$ is unbounded. In the following, we show that the parametric forms of Theorem \ref{th1} are determined by a single parameter.
\begin{Corollary} \label{coroll1}
{Under the case of \eqref{1ge2}, the capacity regions of GS and CS models can be computed as}
\begin{align}
\mathcal{R}_G = \cup_{\alpha \in (0,1]}\Big\{(&R_S,R_J,R_L)\in \mathbb{R}^3_+:~ \nonumber \\
&R_S \leq \frac{1}{2}\log\left(\frac{\alpha\rho^2_2 + 1 - \rho^2_2}{\alpha\rho^2_1 + 1 - \rho^2_1}\right),\nonumber \\
&R_J \geq \frac{1}{2}\log \left(\frac{\alpha\rho^2_1 + 1 - \rho^2_1}{\alpha}\right),\nonumber \\
&R_L \geq \frac{1}{2}\log \left(\frac{\alpha\rho^2_1 + 1 - \rho^2_1}{\alpha(1 - \rho^2_2)}\right)\Big\}, \label{corollary11} \\
\mathcal{R}_C = \cup_{\alpha \in (0,1]}\Big\{(&R_S,R_J,R_L)\in \mathbb{R}^3_+:~ \nonumber \\
&R_S \leq \frac{1}{2}\log\left(\frac{\alpha\rho^2_2 + 1 - \rho^2_2}{\alpha\rho^2_1 + 1 - \rho^2_1}\right),\nonumber \\
&R_J \geq \frac{1}{2}\log \left(\frac{\alpha\rho^2_2 + 1 - \rho^2_2}{\alpha}\right),\nonumber \\
&R_L \geq \frac{1}{2}\log \left(\frac{\alpha\rho^2_1 + 1 - \rho^2_1}{\alpha(1 - \rho^2_2)}\right)\Big\}. \label{corollary12}
\end{align}
{In light of \eqref{1le2}, we have that}
\begin{align}
\mathcal{R}_G = \mathcal{R}_C = \{(&R_S,R_J,R_L):~R_S = 0,~~R_J \geq 0, \nonumber \\
&R_L \geq \frac{1}{2}\log\left(\frac{1}{1-\rho^2_2}\right)\}. \label{corollary13}
\end{align}
\qed
\end{Corollary}

The full proof of Corollary \ref{coroll1} is available in Appendix B.

In Theorem \ref{th1}, one can see that only auxiliary RV $U$ {satisfying the Markov chain $U-X-(Y,Z)$} is present in both regions. A similar conclusion was drawn in \cite{wataoha2010} for {the} secret-key agreement problem, but in the {SKA} systems, it is not trivial whether {the constraint {on} privacy-leakage rate} can be {written} by one auxiliary RV or not. {In this paper}, we have revealed that it is possible to do so. In addition, when the condition \eqref{1le2} is satisfied, the capacity regions of GS and CS models are given in the same form. The optimal values of the secret-key and storage rates are both zero, but that of the privacy-leakage rate can still be positive depending on the joint marginal densities of $(X,Z)$. This is an interesting nature of the SKA systems, which was not seen in the secret-key agreement problems. Even when the encoding procedure is {not needed} ({i.e.,} $U$ is a constant), the {uncontrollable information leaked} to the eavesdropper via the channel $P_{Z|X}$ {is} at minimum rate $I(Z;X)$, corresponding to the capacity of this channel.

As special cases, when $Z$ is independent of other RVs $(\rho_2 = 0)$ and the storage rate is large enough $({R_J} \rightarrow \infty)$, one can easily see that Corollary \ref{coroll1} is reduced to \cite[Theorem 1]{willems2009} and \cite[Theorem 2]{willems2009} for GS and CS models, respectively. Furthermore, for the case where $R_J \ge 0$ and ${R_L} \rightarrow \infty$, let us set $R_J = \frac{1}{2}\log \frac{\alpha\rho^2_1 + 1 - \rho^2_1}{\alpha}$, implying $\alpha = (1-\rho^2_1)/(e^{2R_J}-\rho^2_1)$. Substituting the value of $\alpha$ into the right hand side of $R_S$, with {some} careful {manipulation}, it becomes
\begin{align}
    \frac{1}{2}\log\frac{\Var[Y|X,Z]e^{-2R_J}+\Var[Y|Z](1-e^{-2R_J})}{\Var[Y|X,Z]}, \label{1717}
\end{align}
where $\Var[\cdot]$ denotes the variance of an RV, and $\Var[Y|Z] = \frac{\rho^2_1-\rho^2_2}{\rho^2_1}$, $\Var[Y|X,Z] = \frac{(1-\rho^2_1)(\rho^2_1-\rho^2_2)}{\rho^2_1(1-\rho^2_2)}$. Equation \eqref{1717} is exactly the upper bound {on} the secret-key rate for limited public communication rate ($R_J \ge 0$) derived in \cite[Theorem 4]{wataoha2010}.
\section{Conclusion}
In this paper, we characterized the capacity regions of GS and CS models for Gaussian sources, and showed that only one auxiliary RV was required for expressing the regions. Also, it was demonstrated that the {strong secrecy for} secrecy-leakage of {SKA} systems is achievable by information-spectrum {methods} \cite{han2003}. For future work, we plan to clarify the optimal trade-off for GS and CS models for vector Gaussian sources.

\section*{{Acknowledgement}}
This study was supported in part by JSPS KAKENHI Grant Numbers JP20K04462 and JP18H01438.

\appendices
\section{Proof of Theorem \ref{th1}}
{Due to the space limitations, we only prove GS model under the condition \eqref{1ge2}, which is the most difficult {case} for deriving Theorem \ref{th1}.}
The proof {for} CS model is omitted since it can be done similarly to {GS model} with merely adding an extra procedure; one-time pad operation.
\vspace{0mm}
\subsection{Converse Part}
In \cite{kc2015}, the capacity region of GS model for general discrete sources, denoted by $\mathcal{R}'_G$, is given by
\begin{Theorem} (Kittichokechai and Caire \cite[Theorem 2]{kc2015})
\begin{align}
\mathcal{R}'_G = &\bigcup_{P_{V|U},P_{U|X}}\{(R_S,R_J,R_L) \in \mathbb{R}^3_+ : \nonumber \\
&R_S \leq I(Y;U|V) - I(Z;U|V) ,\nonumber \\
&R_J \geq I(X;U|Y),\nonumber \\
&R_L \geq I(X;U,Y) - I(X;Y|V) + I(X;Z|V)\}, \label{theorem4}
\end{align}
where auxiliary RVs $U,V$ satisfy the Markov chain $V-U-X-(Y,Z)$, and $|\mathcal{V}| \le |\mathcal{X}| + 3$ and $|\mathcal{U}| \le (|\mathcal{X}| + 3)(|\mathcal{X}| + 2)$.
\qed
\end{Theorem}

It can be easily verified that the above result also holds for Gaussian sources. One can see that the bounds of $R_J$ for both Equations \eqref{theorem1} and \eqref{theorem4} remain unchanged, so {they can be shown in the same way}. We need to check that other constraints, i.e., {for} $R_S,R_L$, hold. Consider the case of {\eqref{1ge2}}. For {the} degraded {wiretap} channels, e.g., $V-U-X-Y-Z$, it holds that $I(Y;V|Z) =I(Y;V)-I(Z;V) \ge 0$. Then, we can transform the {bound} on the secret-key rate as
\begin{align}
    R_S & \le I(Y;U|V) - I(Z;U|V) \nonumber \\
    &\overset{\mathrm{(a)}}= I(Y;U) - I(Y;V) - (I(Z;U)-I(Z;V)) \nonumber \\
    &= I(Y;U) - I(Z;U) - (I(Y;V)-I(Z;V)) \nonumber \\
    &\overset{\mathrm{(b)}}= I(Y;U|Z) - I(Y;V|Z) \nonumber \\
    &\le I(Y;U|Z),
\end{align}
where (a) follows by the Markov chains $V-U-Y$ and $V-U-Z$, and (b) is due to $U-Y-Z$ and $V-Y-Z$. For the privacy-leakage rate,
\begin{align}
    R_L &\geq I(X;U,Y) - I(X;Y|V) + I(X;Z|V) \nonumber \\
        &\overset{\mathrm{(c)}}= I(X;U|Y) + I(Y;V) + I(X;Z) - I(Z;V) \nonumber \\
        &= I(X;U|Y) + I(X;Z) + I(Y;V|Z) \nonumber \\
        &\ge I(X;U|Y) + I(X;Z),
\end{align}
{where (c) is due to the Markov chains $V-X-Y$ and $V-X-Z$.}
This wraps up the converse proof.
\qed
\vspace{0mm}
\subsection{Achievability}
Our proof technique is similar to the protocol used in \cite{wataoha2010}. Since the analyses {on the} error probability, secret-key and storage rates, and {strong secrecy for} secrecy-leakage are similar to the arguments discussed in \cite{wataoha2010}, they will be mentioned briefly. However, we will {describe} the {analysis on} the bound of privacy-leakage rate, which was not taken into account in the literature, in {detail}.

Fix the test channel $P_{U|X}$ and let $\gamma$ be small enough positive. Set $R_S = I(Y;U|Z) - 6\gamma$, $R_J = I(X;U|Y) + 4\gamma$, and $R_L = I(X;U|Y) + I(X;Z) + 3\gamma$, and the sizes of {the} set of helpers $|\mathcal{J}| = \exp\{nR_J\}$, and the set of secret keys $|\mathcal{S}| = \exp\{nR_S\}$. Define the sets
\begin{align*}
    \mathcal{T}_n &= \Big\{(u^n,x^n): \frac{1}{n}\log \frac{P_{U^n|X^n}(u^n|x^n)}{P_{U^n}(u^n)} \le I(X;U) + \gamma\Big\}, \nonumber \\
    \mathcal{A}_n &= \Big\{(u^n,y^n): \frac{1}{n}\log \frac{P_{Y^n|U^n}(y^n|u^n)}{P_{Y^n}(y^n)} \ge I(Y;U) - \gamma\Big\}, \nonumber \\
    \mathcal{B}_n &= \Big\{(u^n,x^n,z^n): \nonumber \\
    &~~~~~~\frac{1}{n}\log \frac{P_{X^n|U^n,Z^n}(x^n|u^n,z^n)}{P_{X^n|Z^n}(x^n|z^n)} \ge I(X;U|Z) - \gamma\Big\}.
\end{align*}

Next we specify the codebook, and the enrollment and authentication procedures.

\medskip
\noindent{\em Generation of Codebook $\mathcal{C}_n$}:
Generate $\exp\{n(I(X;U) + 2\gamma)\}$ i.i.d.\ sequences of $\Tilde{u}^n_i, i \in [1: \exp\{n(I(X;U) + 2\gamma)\}]$, from $P_{U}$ and denote the set of these sequences as $\mathcal{Q}_n$. Let $g_n: \mathbb{R} \rightarrow \mathcal{Q}_n\subset \mathbb{R}$ be the quantization function of the biometric source sequence $x^n$ into $\Tilde{u}^n$. The quantization rule of the function $g_n$ is that it looks for a $\Tilde{u}^n_i$ such that $(\Tilde{u}^n_i,x^n) \in \mathcal{T}_n$. In case there are multiple such $\Tilde{u}^n$, the encoder picks one at random. More specially, $\mathcal{Q}_n = \{g_n(x^n)|x^n \in \mathbb{R}^n\}$ and the size $|\mathcal{Q}_n| = \exp\{n(I(X;U) + 2\gamma)\}$. Prepare $M_J = e^{nR_J}$ bins. Randomly assign each $\Tilde{u}^n \in \mathcal{Q}_n$ to one of the bins according to a function $\phi_n: \mathcal{Q}_n \rightarrow \mathcal{J}$. Let $j = \phi_n(\Tilde{u}^n), j \in \mathcal{J},$ denote the index of the bin to which $\Tilde{u}^n$ belongs. A function $f_n: \mathcal{Q}_n \rightarrow \mathcal{S}$ is selected uniformly from $\mathcal{F}_n$ {so that it satisfies} that $P_{F_n}(\{f_n \in \mathcal{F}_n: f_n(\Tilde{u}^n) = f_n({{{\widehat{u}^n}}})\}) \le \frac{1}{|\mathcal{S}|}$, where $P_{F_n}$ is a uniform distribution on $\mathcal{F}_n$, for any distinct sequences $\Tilde{u}^n \in \mathcal{Q}_n$ and ${\widehat{u}^n} \in \mathcal{Q}_n$.

In the encoding and decoding processes, we fix the set $\mathcal{Q}_n$ and the random functions $\phi_n$ and $f_n$.

\medskip
\noindent{\em Encoding}:~~~ Observing $x^n$, the encoder utilizes the function $g_n$ to quantize this sequence to $\Tilde{u}^n$. Then, it computes the bin's index $j = \phi_n(\Tilde{u}^n)$ and generates a secret key $s = f_n(\Tilde{u}^n)$ by a function $f_n: \mathcal{Q}_n \rightarrow \mathcal{S}$, subsequently specified in {Lemmas} \ref{wo} and \ref{naito}. The index $j$ is {shared with the decoder for authentication}. If there is no such $\Tilde{u}^n$, set $(j,s) = (1,1)$. 

\medskip
\noindent{\em Decoding}:~~~Seeing $y^n$ and $j$, the decoder looks for a unique $\hat{u}^n$ such as $j = \phi_n(\hat{u}^n)$ and $(y^n,\hat{u}^n) \in \mathcal{A}_n$. If {such a $\hat{u}^n$ is found}, then the decoder {sets} $\psi_n(y^n,j) = \hat{u}^n$ by a function $\psi_n:\mathcal{J} \times \mathbb{R}^n \rightarrow \mathcal{Q}_n$, and distills the secret key $\hat{s} = f_n(\hat{u}^n)$. Otherwise, the decoder outputs $\hat{s}=1$ and error is declared.

\medskip
The random codebook $\mathcal{C}_n$ consists of the set {$\mathcal{Q}_n = \{\Tilde{U}^n_i : i \in [1: \exp\{n(I(X;U) + 2\gamma)\}]\}$} and the functions $(g_n,\phi_n,\psi_n,f_n)$, and it is revealed to all parties. In the achievability proof, we evaluate the averaged performance of the system, i.e., conditions \eqref{errorp}--\eqref{privacy}, over all possible $\mathcal{C}_n$.

\medskip
Before proceeding to the detailed analysis, we introduce some important lemmas that will be used in the sequel.

\begin{Lemma} (Iwata and Muramatsu \cite[Lemma 11]{iwatamura2002}) \label{wataoha10}
It holds that
\begin{align}
    &\mathbb{E}_{\mathcal{C}_n}[\Pr\{(g_n(X^n),Y^n) \notin \mathcal{A}_n~{\rm or}~(g_n(X^n),{X^n},Z^n) \notin \mathcal{B}_n\}] \nonumber \\
    &\le 2\sqrt{\delta_n}+\Pr\{(U^n,X^n) \in \mathcal{T}_n\} + \exp\{-e^{n\gamma}\}, \label{3939393}
\end{align}
where
\begin{align}
    \delta_n = \Pr\{(U^n,Y^n) \in \mathcal{A}_n~{\rm or}~(U^n,X^n,Z^n) \in \mathcal{B}_n\},
\end{align}
and where $\mathbb{E}_{\mathcal{C}_n}[\cdot]$ denotes the expectation over the random codebook $\mathcal{C}_n$.
\qed
\end{Lemma}
As mentioned in \cite{wataoha2010}, it can be shown that $\Pr\{(U^n,X^n) \notin \mathcal{T}_n\}$ and $\delta_n$ go to zero exponentially by the Chernoff bound.

\begin{Lemma} (Iwata and Muramatsu \cite[Proof of Theorem 1]{iwatamura2002}) \label{iwatamura}
The error probability averaged over the random assemble is bounded by 
\begin{align}
    \mathbb{E}_{\mathcal{C}_n}&[\Pr\{g_n(X^n) \neq \psi_n({\phi_n(g_n(X^n))},Y^n)\}] \nonumber \\
    &\le e^{-\gamma n} + \mathbb{E}_{\mathcal{C}_n}[\Pr\{(g_n(X^n),Y^n) \notin \mathcal{A}_n\}]. \label{iwatamura11}
\end{align}
\qed
\end{Lemma}

Here, we define a security {measure}
\begin{align}
    \mu_n = \int_{z^n}P_{Z^n}(z^n)\|P_{SJ|Z^n=z^n,\mathcal{C}_n}-P_{\Tilde{S}}P_{J|Z^n=z^n,\mathcal{C}_n}\|dz^n,
\end{align}
where {$\|P_A-P_B\|$ and $P_{\Tilde{S}}$ denote the variational distance between probability distributions $P_A$ and $P_B$}, and the uniform distribution on the set $\mathcal{S}$, respectively.

\begin{Lemma} (Watanabe and Oohama \cite[Lemma 12]{wataoha2010}) \label{wo}
An upper bound of the measure $\mu_n$ averaged over the random codebook is given by
\begin{align}
    \mathbb{E}_{\mathcal{C}_n}[\mu_n] \le e^{-\frac{n\gamma}{2}} + \mathbb{E}_{\mathcal{C}_n}[\Pr\{(g_n(X^n),X^n,Z^n) \in \mathcal{B}_n\}]. \label{434343}
\end{align}
\qed
\end{Lemma}

Note that the right-hand side of Equation \eqref{434343} decays {exponentially} since {so does} the second term (cf.\ \eqref{3939393}).

\begin{Lemma} (Naito et al.\ \cite[Lemma 3]{Naito2008}) \label{naito}
We have that
\begin{align}
    H&(S|J,Z^n,\mathcal{C}_n) \nonumber \\
    &\ge (1-\mathbb{E}_{\mathcal{C}_n}[\mu_n])\log M_S + \mathbb{E}_{\mathcal{C}_n}[\mu_n]\log\mathbb{E}_{\mathcal{C}_n}[\mu_n].
\end{align}
\qed
\end{Lemma}

Using Lemma \ref{iwatamura}, the {ensemble average of the} error probability of encoding and decoding can be made exponentially vanishing for large enough $n$. For the analysis of {the} secret-key rate, this can be proved via Lemma \ref{naito}. The bound on the storage rate is straightforward from the rate setting.

\noindent{\em Analysis of Secrecy-Leakage}: We can {expand} the left-hand side of \eqref{secrecy} as
\begin{align}
    I(S;J,Z^n|\mathcal{C}_n) &= H(S|\mathcal{C}_n) - H(S|J,Z^n,\mathcal{C}_n) \nonumber \\
    &\le \log M_S - H(S|J,Z^n,\mathcal{C}_n) \nonumber \\
    &\overset{\rm (a)}\le \mathbb{E}_{\mathcal{C}_n}[\mu_n](\log M_S - \log\mathbb{E}_{\mathcal{C}_n}[\mu_n]) \nonumber \\
    &\le \gamma
\end{align}
{for sufficiently large $n$}, where (a) follows from Lemma \ref{naito} and the last inequality is due to Lemma \ref{wo}.

\noindent{\em Analysis of Privacy-Leakage}: For \eqref{privacy}, we have that
\begin{align}
    I&(X^n;J,Z^n|\mathcal{C}_n) \nonumber \\
    &= I(X^n;J|\mathcal{C}_n) + I(X^n;Z^n|J,\mathcal{C}_n) \nonumber \\
    &\overset{\rm (b)}= H(J|\mathcal{C}_n) + h(Z^n|J,\mathcal{C}_n) - h(Z^n|J,X^n,\mathcal{C}_n) \nonumber \\
    &\overset{\rm (c)}= H(J|\mathcal{C}_n) + h(Z^n|J,\mathcal{C}_n) - h(Z^n|X^n) \nonumber \\
    & \overset{\rm (d)}\le H(J|\mathcal{C}_n) + h(Z^n) - h(Z^n|X^n) \nonumber \\
    &\le n(I(X;U|Y)+4\gamma) + nI(X;Z) \nonumber \\
    &= n(R_L + \gamma).
\end{align}
where (b) holds as $J$ is a function of $X^n$, (c) {holds} because for a given codebook $\mathcal{C}_n$, $J-X^n-Z^n$ forms a Markov chain and {the codebook $\mathcal{C}_n$ is independent of $(X^n,Z^n)$}, and (d) follows because conditioning {does not increase} entropy.

Finally, applying the selection lemma \cite[Lemma 2.2]{BB}, there exists at least one good codebook that satisfies all the conditions in Definition \ref{def1}.
\qed
\section{Proof of Corollary \ref{coroll1}}

In the same manner of Appendix A, we give only the proof of $\mathcal{R}_G$ (cf.\ \eqref{corollary11}) in the case of \eqref{1ge2}. For the achievability part, fix $0 < \alpha \le 1$. Let $U \sim \mathcal{N}(0,1-\alpha)$ and $\Theta \sim \mathcal{N}(0,\alpha)$. Assume that
$
X = U + \Theta.
$
Then, we have that
$
Y = \rho_1U + \rho_1\Theta + N_y
$,
$Z =\rho_2U + \rho_2\Theta + \frac{\rho_2}{\rho_1}N_y + N_z.
$
From these relations, it is not so difficult to see that
\begin{align}
    I(X;U) &= \frac{1}{2}\log(\frac{1}{\alpha}),~~
    I(Y;U) = \frac{1}{2}\log(\frac{1}{\alpha\rho^2_1 + 1 - \rho^2_1}), \nonumber \\
    I(Z;U) &= \frac{1}{2}\log(\frac{1}{\alpha\rho^2_2 + 1 - \rho^2_2}). \label{zu}
\end{align}
Note that due to the Markov chains $U-X-Y-Z$, we can write that
\vspace{0mm}
\begin{align}
    I(Y;U|Z) &= I(Y;U) - I(Z;U), \nonumber \\
    I(X;U|Y) &= I(X;U) - I(Y;U).
\end{align}
Substituting all equations in \eqref{zu} into the right-hand side of \eqref{theorem1}, one can see that any rate tuple contained in the right-hand side of \eqref{corollary11} is achievable.

For the converse part, it is a bit {more} involved. Here, we prove this part by making use of conditional entropy power inequality (EPI) \cite[Lemma II]{bergmans1974}. Note that each constraint in the right-hand side of \eqref{theorem1} can be transformed as
\begin{align}
    R_S &\le I(Y;U|Z) = h(Y|Z) - h(Y|U,Z)  \nonumber \\
    &= \frac{1}{2}\log 2 \pi e (1-\rho^2_2/\rho^2_1)- h(Y|U,Z), \label{iyuuuz} \\
    R_J & \ge I(X;U|Y) \overset{\mathrm{(a)}}= I(X;U|Z) - I(Y;U|Z) \nonumber \\
    &= h(X|Z) - h(X|U,Z) - h(Y|Z) + h(Y|U,Z) \nonumber \\
    &= \frac{1}{2}\log\frac{1-\rho^2_2}{1-\rho^2_2/\rho^2_1} - h(X|U,Z) + h(Y|U,Z),
\end{align}
\begin{align}
    R_L &= h(X|Z) - h(X|U,Z) - h(Y|Z) + h(Y|U,Z) \nonumber \\
    &~~~+ I(X;Z) \nonumber \\
    &\overset{\mathrm{(b)}}\ge \frac{1}{2}\log\frac{1}{1-\rho^2_2/\rho^2_1} - h(X|U,Z) + h(Y|U,Z),\label{ixzuuu}
\end{align}
where (a) is due to the Markov chain $Z-Y-X-U$, and (b) follows from the property that $h(X)=h(Y)$ for Gaussian RVs with unit variances, and thus $ h(X|Z) - h(Y|Z) + I(Z;X) = I(Z;Y) = \frac{1}{2}\log\frac{1}{1-\rho^2_2/\rho^2_1}$.

So as to bound the region $\mathcal{R}_G$, we need to find a lower bound {on} {$h(Y|U,Z)$} for fixed {$h(X|U,Z)$}.

The following fine setting plays {an important} role to {bound} the conditional entropy {$h(Y|U,Z)$}. Now let us set
\begin{align}
    h(X|U,Z) &= \frac{1}{2} \log 2 \pi e \left(\frac{\alpha(1-\rho^2_2)}{\alpha\rho^2_2+1-\rho^2_2}\right) \label{setting111}
\end{align}
for $0 < \alpha \le 1$. This setting comes from the fact that $h(X|U,Z) \le h(X|Z) = \frac{1}{2}\log2\pi e(1-\rho^2_2)$. {Here}, $\alpha = 0$ is excluded since the right hand-side of \eqref{setting111} will go to infinity, but this value is impossible to achieve by Gaussian RVs with finite {variances}, which {are} always assumed in the analysis of Gaussian sources.

In the direction from $X$ to $Y$, using the conditional EPI \cite[Lemma II]{bergmans1974}, we have that
\begin{align}
&e^{2h(Y|U,Z)} \ge e^{2h(\rho_1X|U,Z)} + e^{2h(N_y|U,Z)} \nonumber \\
&~~= \rho^2_1 e^{2h(X|U,Z)} + e^{2h(N_y)} \nonumber \\
&~~\overset{\mathrm{(c)}}= 2 \pi e \left(\frac{\alpha \rho^2_1(1-\rho^2_2)}{\alpha\rho^2_2+1-\rho^2_2}\right) + 2 \pi e(1-\rho^2_1) \nonumber \\
&~~= 2 \pi e \left(\frac{\alpha \rho^2_1(1-\rho^2_2) + (1-\rho^2_1)(\alpha\rho^2_2+1-\rho^2_2)}{\alpha\rho^2_2+1-\rho^2_2}\right),\label{hxuz}
\end{align}
where (c) follows {from} \eqref{setting111}.
Now let us focus only on the numerator of \eqref{hxuz} (inside the biggest parenthesis). We continue scrutinizing it as
\vspace{0mm}
\begin{align}
    \alpha &\rho^2_1(1-\rho^2_2) + (1-\rho^2_1)(\alpha\rho^2_2+1-\rho^2_2) \nonumber \\
    &=\alpha \rho^2_1 -2\alpha\rho^2_1\rho^2_2 + \alpha\rho^2_2 + 1 -\rho^2_2 -  \rho^2_1 + \rho^2_1\rho^2_2 \nonumber \\
    &= (\alpha \rho^2_1 + 1 - \rho^2_1)(1-\frac{\rho^2_2}{\rho^2_1}) + \frac{\rho^2_2}{\rho^2_1}(\alpha \rho^2_1 + 1 - \rho^2_1) \nonumber \\
    &~~~+\rho^2_2(-2\alpha\rho^2_1 -1 + \alpha + \rho^2_1) \nonumber \\
    &= (\alpha \rho^2_1 + 1 - \rho^2_1)(1-\frac{\rho^2_2}{\rho^2_1})+ \frac{\rho^2_2}{\rho^2_1}(1-\rho^2_1)(2\alpha\rho^2_1+1-\rho^2_1) \nonumber \\
    &\overset{\mathrm{(d)}}\ge (\alpha \rho^2_1 + 1 - \rho^2_1)(1-\frac{\rho^2_2}{\rho^2_1}), \label{numerator1}
\end{align}
where (d) follows because $\alpha \in (0,1]$ and $\rho^2_1,\rho^2_1 < 1$.
{Plugging} \eqref{numerator1} into \eqref{hxuz}, we obtain that
\begin{align}
    e^{2h(Y|U,Z)} &\ge 2 \pi e \left(\frac{(\alpha \rho^2_1 + 1 - \rho^2_1)(1-\rho^2_2/\rho^2_1)}{\alpha\rho^2_2+1-\rho^2_2}\right).
\end{align}
Therefore,
\begin{align}
    h(Y|U,Z) \ge \frac{1}{2}\log 2 \pi e \left(\frac{(\alpha \rho^2_1 + 1 - \rho^2_1)(1-\rho^2_2/\rho^2_1)}{\alpha\rho^2_2+1-\rho^2_2}\right). \label{hyuzzzz}
\end{align}

Finally, substituting \eqref{setting111} and \eqref{hyuzzzz} into \eqref{iyuuuz}--\eqref{ixzuuu}, the converse proof of the region $\mathcal{R}_G$ is completed.
\qed

\end{document}